\documentclass[aps, prd, reprint, superscriptaddress, nofootinbib, bibnotes]{revtex4-2}

\usepackage{graphicx, subfigure}        
\usepackage{aas_macros} 
\usepackage{amssymb,amsmath,placeins,epsfig,array,bm}
\usepackage[usenames,dvipsnames]{color}
\usepackage[normalem]{ulem}
\usepackage[breaklinks,colorlinks,urlcolor=blue,citecolor=blue,linkcolor=blue]{hyperref}
\usepackage{float}
\usepackage[T1]{fontenc}
\usepackage{times}
\usepackage{soul}
\usepackage[capitalise]{cleveref}
\newcommand{\IVEi}{IVE$_{\mathrm{virial}}$}
\newcommand{\IVEo}{IVE$_{\mathrm{infall}}$}
\newcommand{\rvir}{$r_\mathrm{200m}$}
\newcommand{\bol}[1]{\boldsymbol{#1}}

\begin{document}

\title{Explaining dark matter halo density profiles with neural networks}

\author{Luisa Lucie-Smith}
\email[]{luisals@mpa-garching.mpg.de}
\affiliation{Max-Planck-Institut f{\"u}r Astrophysik, Karl-Schwarzschild-Str. 1, 85748 Garching, Germany}

\author{Hiranya V. Peiris}
\affiliation{Department of Physics \& Astronomy, University College London, Gower Street, London WC1E 6BT, UK}
\affiliation{The Oskar Klein Centre for Cosmoparticle Physics, Stockholm University, AlbaNova, Stockholm, SE-106 91, Sweden}

\author{Andrew Pontzen}
\affiliation{Department of Physics \& Astronomy, University College London, Gower Street, London WC1E 6BT, UK}

\date{\today}

\begin{abstract}
We use explainable neural networks to connect the evolutionary history of dark matter halos with their density profiles. The network captures independent factors of variation in the density profiles within a low-dimensional representation, which we physically interpret using mutual information. Without any prior knowledge of the halos' evolution, the network recovers the known relation between the early time assembly and the inner profile, and discovers that the profile beyond the virial radius is described by a single parameter capturing the most recent mass accretion rate. The results illustrate the potential for machine-assisted scientific discovery in complicated astrophysical datasets.
\end{abstract}


\maketitle

\textbf{Introduction} -- 
In the modern picture of structure formation, galaxies form at the center of extended, overdense `halos' of dark matter, which originate from small fluctuations in the density of matter in the early Universe and undergo highly non-linear dynamical processes throughout their evolution \cite{Bardeen1986, Blumenthal1984, Davis1985, Frenk1988, White1991}. The history of a halo determines its final structure, commonly parametrized by the spherically-averaged radial density profile. Halo density profiles are not only key ingredients of the galaxy-halo connection in cosmological analyses and of direct and indirect dark matter searches; they are also powerful observational testbeds of fundamental physics. This is because their shape, from the inner core to the outskirts in the proximity of the splashback radius, is sensitive to the nature of dark matter and modifications to gravity \cite{Bechtol2022, Drlica-Wagner2019}. 

Observationally, it has recently become possible to measure weak lensing and 3D density profiles through a combination of multi-wavelength data; upcoming data from \emph{Euclid}, Rubin, and DESI will provide even more detailed measurements of the density profiles of halos from clusters to dwarf galaxies \cite{LEAUTHAUD2020, Ivezic2018, Mandelbaum2018}. Achieving the potential impact of these measurements requires determining the physical effects that control the shape of the density profiles. However, current theoretical models are limited to empirical fitting functions such as the Navarro–Frenk–White (NFW) \cite{Navarro1997} and Einasto \cite{Einasto1965} profiles; these do not explain the physical origin of the profiles' universal shape seen in numerical simulations \cite{Navarro2004, Wang2020}. Understanding the connection between the formation history and the density profile also offers the possibility of using observational constraints on the latter to estimate the halos' mass accretion rate. This will yield valuable constraints on galaxy formation \cite{Moster2018, Behroozi2019, WechslerTinker2018, Hearin2021} as well as extensions to the cosmological model.

In this work, we use an {\it explainable AI} framework to connect the formation histories of dark matter halos to their density profiles. Our goal differs from typical uses of machine learning in cosmology, such as emulating the output of computationally expensive simulations \cite{McClintock2019, Jamieson2023} or accelerating the estimation of cosmological parameters from data \cite{Kacprzak2022, Kreisch2022, Wang2023, Akhmetzhanova_2023}. In Ref.~\cite{LucieSmith2022}, we used an interpretable deep learning framework to build a new model for the spherically-averaged halo density profiles, generalizing over existing empirical fitting functions. The framework, which we denoted an \textit{interpretable variational encoder} (IVE), was trained to capture all the information used by the neural network to predict the profile, given the 3D density field around the halo center, within a compact, low-dimensional latent representation.
We require the representation to be disentangled i.e., each latent component captures different, independent factors of variation in the profiles; the latent representation is equivalent to the profiles' degrees of freedom. We found that three components are required (and sufficient) for modeling the profiles out to the halo outskirts: these three components describe respectively the normalization of the profile, and its shape within and beyond the virial radius.

In this \textit{Letter}, we turn to the physical interpretation of the learnt IVE latent representation to investigate how halo density profiles are determined from the halos' formation histories. Although the network was trained only on the present-day density field, we explore whether the latent parameters carry memory of the evolution history of the halos. We measure the information encoded within each latent about the halos' evolution history using the information-theoretic measure of {\it mutual information} (MI). By this metric, the IVE representation and the NFW parametrization similarly highlight a dependence of the profile on physical accretion history. However the IVE additionally allows us to measure the connection between a halo's recent evolution history and the density in its far outskirts, something that the NFW profile does not capture.

\vspace{0.3cm}
\textbf{Background} -- 
We begin by briefly reviewing the current understanding of the physics of halo density profiles. The NFW profile is the most widely used fitting function for the halo density profile. It is given by
\begin{equation}
\rho (r) = \frac{\rho_s}{r/r_s \left( 1 + r/r_s \right)^2} \, ,
\label{eq:nfw}
\end{equation}
where $r_s$ and $\rho_s$ are the scale radius and characteristic density, respectively. The scale radius is often re-written in terms of a concentration parameter $c \equiv r_\mathrm{200m}/r_s$, so that the NFW profile depends on the virial radius $r_\mathrm{200m}$ and concentration $c$. The virial radius $r_\mathrm{200m}$ is typically adopted as a proxy for the halo boundary, and defined as the radius which contains a mean density that is 200 times the mean density of the Universe. High-resolution simulations have revealed this functional form to be `universal': it provides a good fit to stacked profiles of halos for a large range of halo masses \cite{Wang2020, Ludlow2014}, for several different cosmological models \cite{Diemer2015, Correa2015, Ludlow2016, Prada2012, Brown2021}, and even in the absence of hierarchical growth \cite{Huss1999, Wang2009, Moore1999}. This suggests that universal density profiles are a generic feature that arises from collisionless gravitational collapse.

Despite the lack of a first-principles explanation for the self-similarity of halo density profiles, some insights have been gained from studying the correlation between the NFW concentration and summary statistics of the halo evolution process. Mass, concentration and halo formation time all correlate: on average, low-mass halos assemble earlier and have higher characteristic densities (or concentration), reflecting the larger background density at earlier times \cite{Navarro1997, Bullock2001, Wechsler2002, Prada2012, Ludlow2014, Zhao2009, dalal2010}. This  description can explain the qualitative trend of the mean concentration as a function of halo mass, but not the large residual scatter in concentration seen in simulations \cite{Wechsler2002, Rey2019}. It is also limited to the simplest summary statistic of the halo evolution history, i.e. the halo formation time. The scatter in concentration at fixed mass has been shown to be at least in part connected to merger events during the halo assembly process \cite{Roth2015, Rey2019, WangK2020}. Further work has suggested that the self-similarity of halos may be related to the self-similarity of the halo mass assembly history \cite{Ludlow2014}, although this has only been validated on stacked profiles of well-behaved,`relaxed' halos.

The situation worsens when modeling profiles beyond the virial radius: the halo outskirts strongly deviate from the NFW form due to the presence of the splashback radius, where particles reach the apocenter of their first orbit. Recent work has focused on modeling the location of the splashback radius, finding that it is sensitive to the late-time mass accretion rate \cite{More2015, Adhikari2014, Shi2016, Adhikari2018, Diemer2017, Shin2023}. Modeling the full shape of the outer profile remains a difficult task due to its intrinsically non-equilibrium nature, leading to a reliance on multi-parameter fitting functions with little physical explainability \cite{Diemer2014, Diemer2022}.

\begin{figure}
      \centering
        \includegraphics[width=\columnwidth]{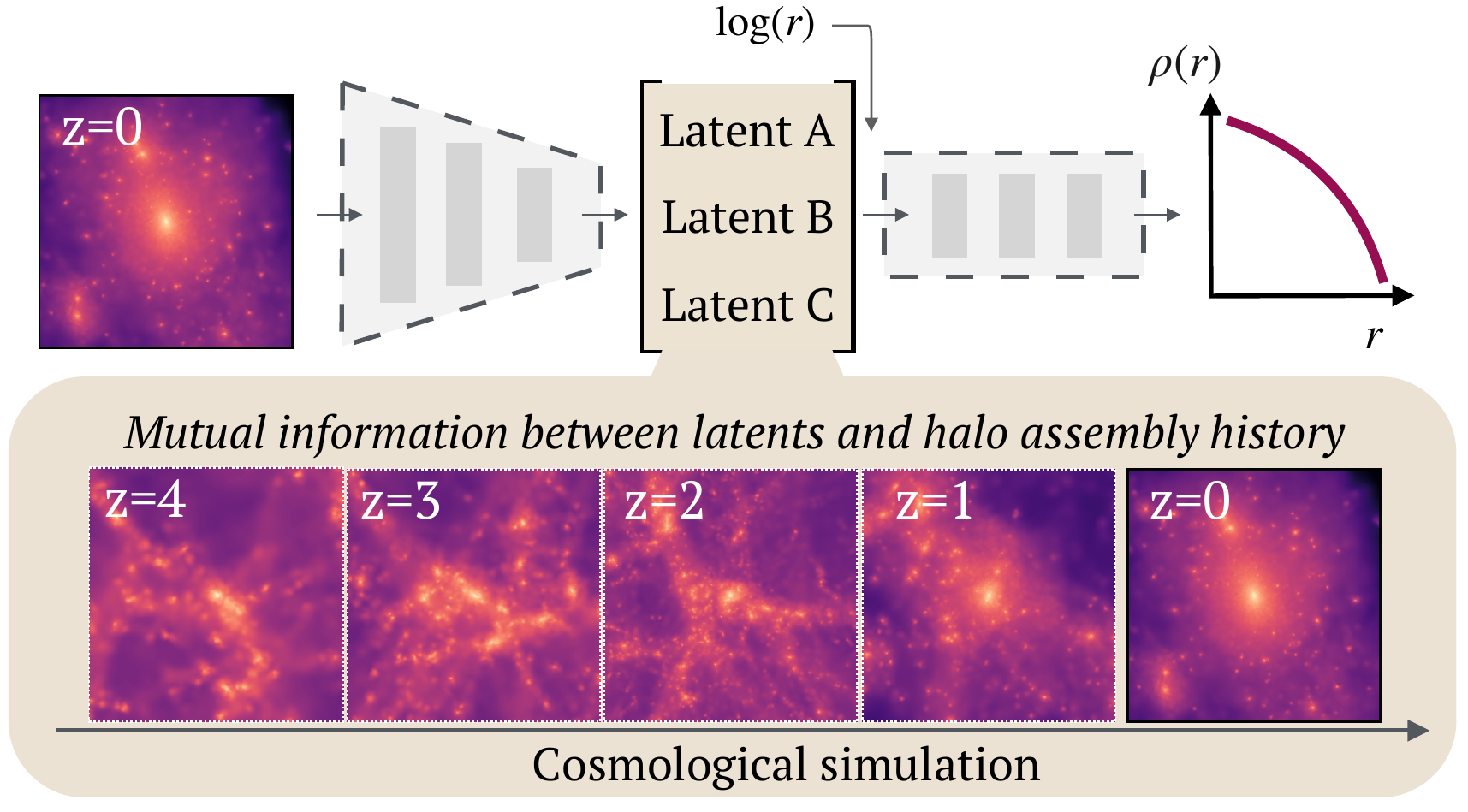}
        \caption{A neural network is trained to discover the underlying degrees of freedom in halo density profiles in the form of a latent representation, when presented with the full 3D density structure of a halo. We physically interpret the discovered representation by measuring the MI between the latent parameters and the assembly history of the halos.}
    \label{fig:illustration}
\end{figure}

\vspace{0.3cm}
\textbf{Deep learning model} -- 
The IVE architecture used in this work has two main components: the encoder, mapping the 3D density field to a low-dimensional latent representation, and the decoder, mapping the latent representation and the query radius $\log(r)$ to the output profile $\log[\rho(r)]$. By design, all the information used by the model to predict the density profiles is captured within the latent representation. An illustration of the model is shown in the top-half of Fig.~\ref{fig:illustration}. The encoder is a 3D convolutional neural network (CNN) with parameters $\phi$ that maps the inputs $\bol{x}$ to a multivariate distribution in the latent space $p_{\phi}(\bol{z} | \bol{x})$. We choose the latent representation to be a set of independent Gaussians, $p_{\phi} (\bol{z} | \bol{x}) = { \prod_{i=1}^L}  \mathcal{N}(\mu_i(\bol{x}), \sigma_i(\bol{x}))$, where $L$ is the dimensionality of the latent space; under this assumption, the encoder maps the inputs $\bol{x}$ to the vectors $\mu = {\mu_i, .., \mu_L}$ and $\sigma = {\sigma_i, .., \sigma_L}$. The decoder of the IVE consists of another neural network model with parameters $\theta$ that maps a sampled latent vector $z \sim p_{\phi} (\bol{z} | \bol{x})$ and a value of the query $\log(r)$ to a single predicted estimate for $\log[\rho_\mathrm{pred}(r)]$. 

A crucial aspect of the IVE that makes the latent space interpretable is that it is \textit{disentangled}: independent factors of variation in the density profiles are captured by different, independent latents. This is achieved through the design of a loss function that minimizes the mean squared error between predicted and ground-truth profiles, while simultaneously maximizing the degree of independence between the latent variables by encouraging those to be as close as possible to independent Gaussians of mean 0 and variance 1 \cite{Higgins2017}. More details on the encoder and decoder architectures and the loss function are presented in Ref.~\cite{LucieSmith2022}.

\vspace{0.3cm}
\textbf{Methods} -- 
We generated the training data from four dark-matter-only $N$-body simulations produced with GADGET-4 \citep{gadget4}, each containing $512^3$ particles in a $(50 \, \mathrm{Mpc} \, h^{-1})^3$  box.
We trained two IVE models for different tasks: one (\IVEi{}) was trained to model the density profile up to the halo virial radius \rvir{}, and the second (\IVEo{}) was trained to model profiles beyond the halo boundary out to $2\,$\rvir{}. The former is used for direct comparison with the NFW profile, which is also designed to model the profile out to the virial radius, and the latter is used to investigate the less studied halo outer profile. The innermost radius of the profiles we consider is $r_\mathrm{min} = 3\, \epsilon$, where $\epsilon$ is the gravitational softening of the simulation; this choice ensures that we can robustly trust the inner profile. The inputs are given by the 3D density field within a $N=131^3$ sub-box of size $L_\mathrm{sub-box} = 0.4 \, \mathrm{Mpc} \, h^{-1}$ for the \IVEi{} model, and of size $L_\mathrm{sub-box} = 0.6 \, \mathrm{Mpc} \, h^{-1}$ for the \IVEo{} one. We considered halos with $\log_{10} ( M/M_{\odot} ) \in [11, 13]$, but for the \IVEo{} model, we further restricted our analysis to halos with \rvir $\leq 150\, \mathrm{kpc}  \, h^{-1}$. These cuts yielded $\sim$17,000 (13,000) halos for training the \IVEi{} (\IVEo{}) model. Further discussion on the training data of the \IVEi{} and \IVEo{} models is presented in Ref.~\cite{LucieSmith2022}. To compare the \IVEi{} results with NFW, we fitted the NFW formula in Eq.~\eqref{eq:nfw} to each halo's density profile using least-squares minimization, and recovered the best-fitting parameters $r_s$ and $\rho_s$. The concentration was then derived using $c=$\rvir{}$/r_s$. A description of the simulations used for training and testing the IVE models can be found in the Supplemental Material.

The first step of the analysis was to verify that the IVE models learn to predict the density profiles at $\sim\,$5\% accuracy, comparable to the accuracy of NFW fits (see Supplemental Material). Crucially, only the $z=0$ snapshots were used for training the IVE models to construct the latent representations mapping the 3D density field to dark matter halo profiles -- i.e., the model had no access to the merger histories of the halos during training. The resulting disentangled latent space directly corresponds to the underlying degrees of freedom in the halo density profiles. Following recent works \cite{Pandey2017, Sarkar2020, LucieSmith2022, Sedaghat2021, Piras2022}, we then used the MI to (i) quantify the information captured by the latent space about the halo density profiles and (ii) connect the IVE latents to the halo's evolution history, showing how the latter determines the present-day density profile. 

The MI was estimated using \textsc{GMM-MI} \cite{Piras2022}, which performs density estimation using Gaussian mixtures and provides  MI uncertainties through bootstrap. Background details on MI are provided in the Supplemental Material. We first measured the MI between each latent and the density profile $\rho(r)$; this allows us to directly link each latent to a degree of freedom in the profile that affects its shape over a certain radial range. We then measured the MI between each latent and the mass assembly history of each halo. This in turn allowed us to connect each degree of freedom describing the density profile of the halo directly to characteristics of the halos' evolution which determines that component.

\begin{figure}
      \centering
        \includegraphics[width=\columnwidth]{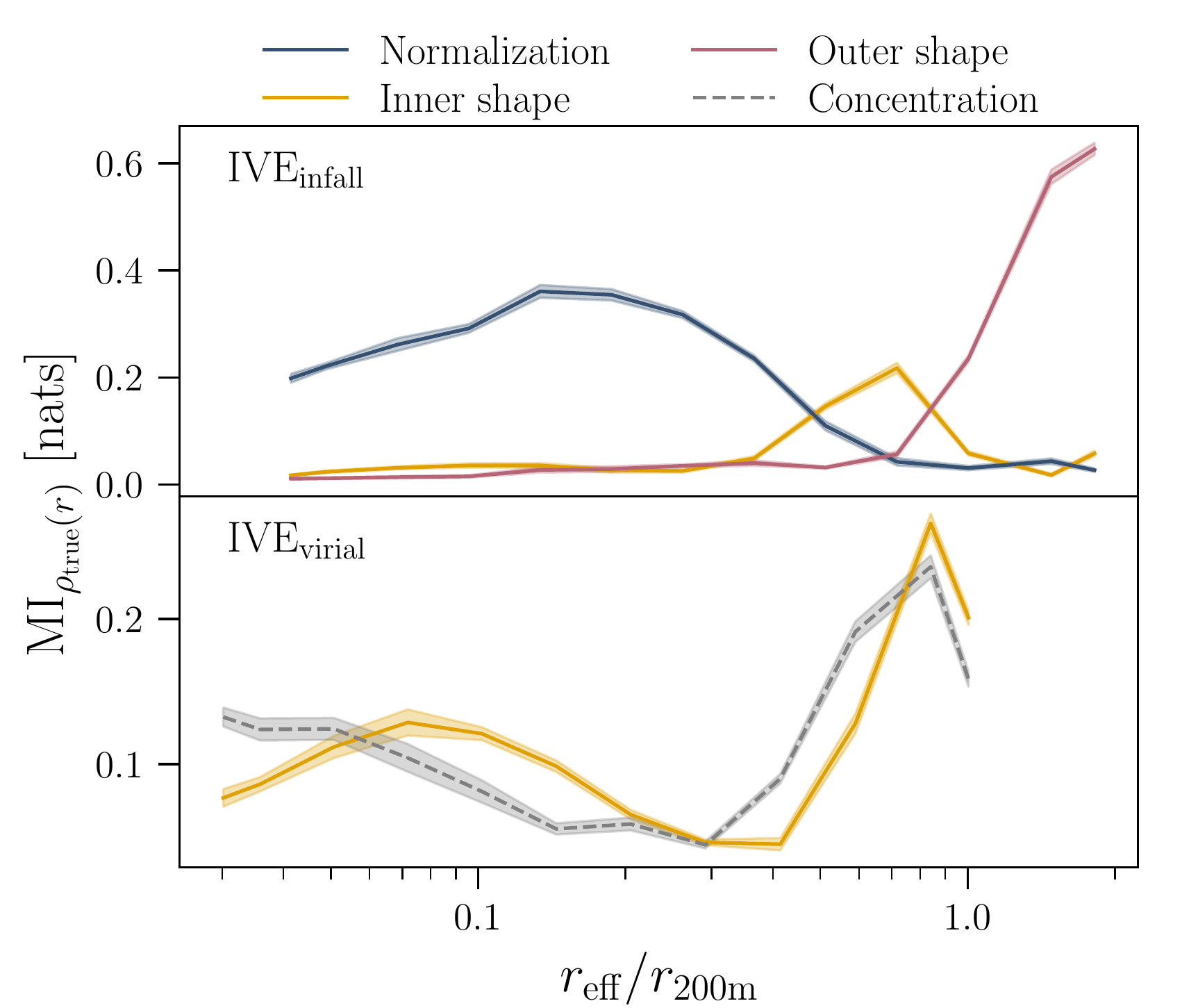}
        \caption{The MI between the latent parameters and the ground-truth halo profiles $\rho_\mathrm{true}(r)$ for the \IVEo{} (\textit{top}) and the \IVEi{} (\textit{bottom}) models. In the \IVEi{} case, we also show MI with the NFW concentration. (For clarity we do not show the \IVEi{} normalization latent, since it behaves identically to the \IVEo{} normalization latent.)}
    \label{fig:MI_latents_truth}
\end{figure}

\vspace{0.3cm}
\textbf{Results} -- 
Figure~\ref{fig:MI_latents_truth} quantifies the information contained within the latents of the \IVEo{} (top panel) and the \IVEi{} (bottom panel) models about the ground-truth density profiles\footnote{We verify the conclusions of our previous work in Ref.~\cite{LucieSmith2022} at higher precision using the new \textsc{GMM-MI} estimator \cite{Piras2022}.}. We show the MI between each latent parameter and the ground-truth profiles, which we denote as MI$_{\rho_\mathrm{true}(r)}$. The three latents discovered by the \IVEo{} describe (i) the normalization of the profile, which dominates the variation in the profiles out $\sim$ \rvir{}$/2$, (ii) the shape of the inner profile, which becomes informative on radial scales approaching  \rvir{}, and (iii) the shape of the outer profile beyond  \rvir{}.
The first two are analogous to the two NFW parameters, mass and concentration, respectively. 
A closer comparison between the inner shape latent of the \IVEi{} model and concentration (bottom panel of Fig.~\ref{fig:MI_latents_truth}) shows that both parameters carry information about the density in the core and on radial scales close to \rvir{}\footnote{The MI between the density profile and the inner latents of the \IVEo{} and \IVEi{} models (in yellow, Fig.~\ref{fig:MI_latents_truth} top and bottom panels, respectively) is 
qualitatively similar, despite the models' different training sets.}. 
This bimodality is due to a compensation effect between the density in the inner region and that close to the virial boundary: at fixed normalization, halos with denser cores become less dense in the outskirts, and vice versa. The MI$_{\rho_\mathrm{true}(r)}$ of the inner latent is shifted towards larger radii compared to that of concentration, suggesting that the former is sensitive to variations in the shape of the profile on larger radial scales than the latter; this distinction will become relevant when physically interpreting the latent and comparing it to concentration.

\begin{figure}
      \centering
        \includegraphics[width=\columnwidth]{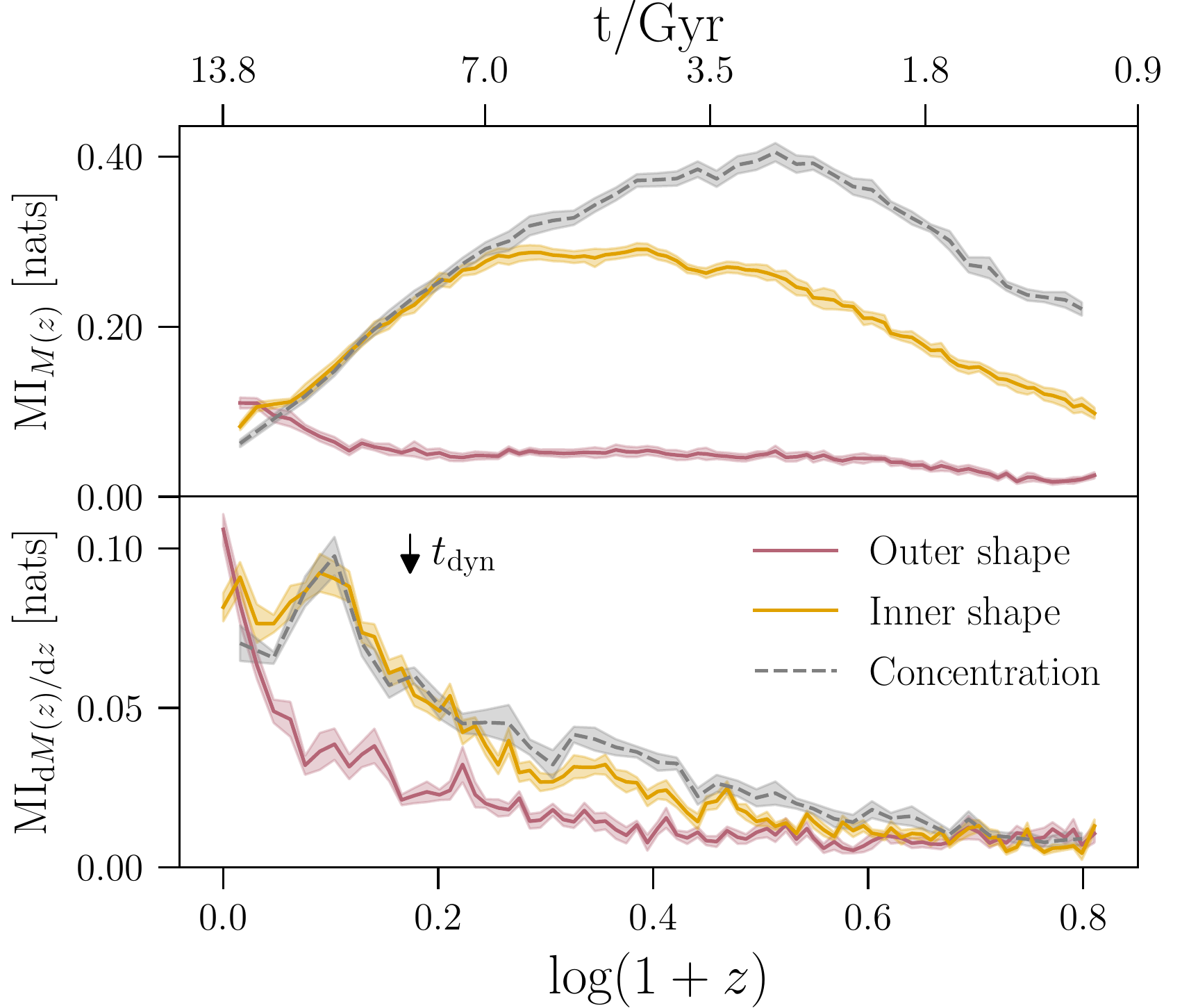}
        \caption{The MI between the latent parameters and the mass accretion histories  (denoted MI$_{M(z)}$; top row), and that between the latent parameters and the mass accretion rate (denoted MI$_{\mathrm{d}M(z)/\mathrm{d}z}$; bottom row). The inner shape latent and the NFW concentration carry memory of the early-time mass assembly history, as well as the later-time mass accretion rate. The outer shape latent carries information about the halos' most recent mass accretion rate over the past dynamical time (indicated by the arrow).}
    \label{fig:MI_latents_MAH}
\end{figure}

We now move on to a physical interpretation of the latents in relation to characteristics of the halos' evolution histories. Recall that the network did not have access to this information during training. The interpretation of the normalization latent is straightforward: it captures the $z=0$ mass of the halo, $M_\mathrm{200m}$. Their MI is $\sim2.07 \pm 0.01$ nats, implying a strong correlation between the two. This also matches expectations from the literature \cite{Navarro1997, Einasto1965}, as halo mass also controls the normalization in the NFW and Einasto fitting functions. To physically interpret the inner and outer shape latents, we measure their MI with two quantities that describe the assembly history of the halos over cosmic time. The first is the mass accretion history, $M_{\mathrm{200m}}(z)/M_{\mathrm{200m}}(z=0)$, which describes the evolution of the halo mass as a function of time $M_{\mathrm{200m}}(z)$ normalized to the present-day halo mass $M_{\mathrm{200m}}(z=0)$. The second is the mass accretion rate $\Gamma(t) \equiv \Delta \ln M_\mathrm{200m}(a)/ \Delta \ln a$ \cite{Diemer2017}, which describes the rate of change in halo mass with respect to the scale factor $a(t)$. The value of the accretion rate depends on the time interval used to compute the change in mass and scale factor; we compute $\Gamma(t)$ by taking the finite difference of the halo masses at each consecutive timestep in the simulation.

Figure~\ref{fig:MI_latents_MAH} shows the MI between the latents and mass accretion history (MI$_{M(z)}$; top row) and that between the latents and the mass accretion rate (MI$_{\mathrm{d}M(z)/\mathrm{d}z}$; bottom row). We first focus on the inner shape latent, which we compare to the NFW concentration. The MI$_{M(z)}$ of the inner shape latent increases with time during the early formation period, peaks at $z\sim 1$, and declines rapidly towards $z=0$; recall that this is the MI with the mass assembly history normalized to the present-day halo mass. This result reveals that the inner shape latent is sensitive to the early assembly history of halos. The MI$_{dM(z)/dz}$ of the same latent reveals that the latter is also sensitive to the later time mass accretion rate. This dual dependence explains the bimodal shape of the MI between the inner latent and the profile (Fig.~\ref{fig:MI_latents_truth}, bottom panel): the early assembly phase determines the shape of the profile in the innermost region of the halo, while the later time mass accretion rate determines the shape of the profile close to the virial radius. We further validate this interpretation in the Supplemental Material.

The NFW concentration shows a similar picture to the inner shape latent. However, its MI$_{M(z)}$ peaks at earlier times ($z\sim0.55$) compared to the inner shape latent. This implies that the inner shape latent carries information about the build up of mass onto the halo over a longer period of time than concentration, which therefore affects the inner halo structure (and the profile) out to larger scales. The sensitivity of the inner latent to later times/larger scales in the profile explains why the inner latent MI$_{\rho_\mathrm{true}(r)}$ is shifted towards larger radial scales than that of concentration (Fig.~\ref{fig:MI_latents_truth}). Moreover, the absolute magnitude of the concentration MI$_{M(z)}$ is higher than that of the inner shape latent; this is because the closer to the halo core the stronger the correlation with the early assembly history due to halos accreting mass `inside-out'. As a result, concentration, which is sensitive to the profile on smaller $r$ than the latent, has a higher MI with the early assembly history than the latent. Finally, the NFW concentration is related to the later time mass accretion rate in a similar way to the inner latent.

Figure~\ref{fig:MI_latents_MAH} shows that the outer shape latent shares information about the mass accretion rate over the past $\sim 5$ Gyr, since this is the period over which the MI roughly doubles (see bottom panel). This timescale corresponds to the halo dynamical time, $t_\mathrm{dyn} \equiv 2\times r_\mathrm{200m}/v_\mathrm{200m}$, defined as the time it takes for material to cross the halo at a typical virial velocity $v_\mathrm{200m} = \sqrt{\mathrm{G}M_\mathrm{200m}/r_\mathrm{200m}}$. This suggests that the outer profile is primarily determined by the infall of dynamically unrelaxed material within the last dynamical time, which has not yet virialized within the halo.

\vspace{0.3cm}
\textbf{Discussion} -- 
Our results show that the IVE framework has extracted a direct connection between the assembly history of cold dark matter halos and their density profiles, without having access to explicit information about the time evolution of the halos during training. 
This has deep implications for understanding the origin of universality in dark matter halos; the universality in the profiles, captured by three degrees of freedom alone, may originate from a universality in the halo assembly histories themselves, since the latents contain comparable amounts of information about both quantities.

Previous work \cite{Ludlow2014} found a resemblance between the shape of the average mass accretion history, expressed in terms of the critical density of the Universe, and the average enclosed mass profile, expressed in terms of its enclosed density, for a selected set of `well-behaved’ halos of similar mass. In the halo outskirts, the profile has been linked to the dynamical accretion history of the halos primarily through the relation between the splashback radius and the mass accretion rate \cite{Adhikari2014, Diemer2014}; existing models make use of multi-parameter fitting functions to capture the dynamical impact on the outer profile \cite{Diemer2022}. 

By contrast, within the IVE framework, the connection between the density profiles and the {\it entire} mass accretion history or mass accretion rate is clearly elucidated through MI. This result was obtained using all halos in the simulations, without requiring a curated sample of well-behaved halos. The IVE rediscovers the known correlation between the inner profile and halo formation time \cite{Bullock2001, Wechsler2002}; it then additionally demonstrates that the complexity of the dynamical, infalling material is encoded in only a single degree of freedom that captures the recent mass accretion rate. In future work, we will use the connection between assembly history, latents, and density profile captured by the IVE framework to build a model that can determine mass accretion histories from density profiles.

In future work we will explore extensions to this work using hydrodynamical simulations. We expect the same IVE framework to successfully disentangle the relevant factors in the baryonic case. Previous work found that baryons primarily impact the inner profile \cite{NavarroEke1996, Duffy2010, PontzenGovernato2014}, and that the results can still be encoded with minimal modifications to the pure dark matter expectations \cite{DiCintio2014, Henson2017}. Conversely, the splashback radius in the halo outskirts remains unchanged when comparing hydrodynamical and dark-matter-only simulations \cite{Aung2021}. Thus, an IVE with the same dimensionality or a single additional dimension should suffice to account for the impact of baryonic physics on the halo profiles for the baryonic feedback models included in the training set simulations.

More broadly, our results represent progress towards enabling \textit{new} machine-assisted scientific discoveries, going beyond artificial rediscovery of known physical laws \cite{Iten2020, Seif2021, Udrescu2020}. Our IVE approach towards this goal consisted of compressing the information within a dataset into a set of minimal ingredients which disentangles the independent factors of variation in the output (interpretability), and can be explained in terms of the physics it represents through MI (explainability). 
The approach shows promise for gaining insight into other emergent properties of the cosmic large-scale structure (e.g. void density profiles \cite{Hamaus2014} and the halo mass function \cite{Tinker2008}), building physical explanations that are more accurate and complete than traditional methods have achieved.

\nocite{SM_citation}
\nocite{Aghanim2020}
\nocite{Stopyra2021}
\nocite{gadget2}
\nocite{Pontzen2018}

\vspace{0.3cm}
\textbf{Acknowledgments} -- 
LLS thanks Benedikt Diemer, Eiichiro Komatsu and Simon White for useful discussions. This project has received funding from the European Research Council (ERC) under the European Union’s Horizon 2020 research and innovation programme (grant agreement nos. 818085 GMGalaxies and 101018897 CosmicExplorer). This work has been enabled by support from the research project grant ‘Understanding the Dynamic Universe’ funded by the Knut and Alice Wallenberg Foundation under Dnr KAW 2018.0067. The work of HVP was additionally supported by the G\"{o}ran Gustafsson Foundation for Research in Natural Sciences and Medicine. AP was additionally supported by the Royal Society. HVP and LLS acknowledge the hospitality of the Aspen Center for Physics, which is supported by National Science Foundation grant PHY-1607611. The participation of HVP and LLS at the Aspen Center for Physics was supported by the Simons Foundation. This work was partially enabled by the UCL Cosmoparticle Initiative.

The contributions from the authors are listed below: \textbf{L.L.-S.}: conceptualization; formal analysis; investigation; methodology; software; validation; visualization; writing - original draft, review \& editing. \textbf{H.V.P.}: conceptualization; methodology; interpretation \& validation; writing - review \& editing. \textbf{A.P.}: methodology; writing - review \& editing.

\setcounter{equation}{0}
\setcounter{figure}{0}
\renewcommand{\theequation}{S.\arabic{equation}}
\renewcommand{\thefigure}{S.\arabic{figure}}
\renewcommand{\thesection}{S.\arabic{section}}

\section*{Supplemental Material}

\subsection{Mutual information}
Mutual information (MI) is a well-established information-theoretic measure of the correlation between two random variables. In contrast to linear correlation measures such as the $r$-correlation, MI captures the full (linear and non-linear) dependence between two variables. In other words, the MI between two variables is zero if and only if these are statistically independent. Additionally, MI is invariant under invertible, non-linear transformations, e.g. translations, rotations and any transformation preserving the order of the original elements. 

Mathematically, the MI between two continuous random variables $x$ and $y$ is given by
\begin{equation}
\mathrm{MI}(x, y) = \int \int p(x, y) \log \left[ \frac{p(x, y)}{p(x) p(y)} \right] dx dy,
\label{eq:MI}
\end{equation}
where $p(x, y)$ is the joint probability density distribution between $x$ and $y$, and $p(x)$ and $p(y)$ are their marginal distributions, respectively. The $\log$ is the natural logarithm, so that MI is computed in natural units of information (nats). 

The main challenge in estimating MI is obtaining a density estimate for the joint probability distribution $p(x, y)$ in Eq.~\eqref{eq:MI}. We make use of the publicly-available software GMM-MI \cite{Piras2022}, which performs density estimation using Gaussian mixtures and additionally provides MI uncertainties through bootstrap.

\subsection{Simulations}
We begin with a description of the simulations used for training the IVE models and subsequent interpretation of the learned representation. We ran four dark-matter-only $N$-body cosmological simulations using the publicly-available code GADGET-4 \citep{gadget4} assuming a \textit{Planck} $\Lambda$CDM cosmological model \citep{Aghanim2020}. We evolved $N=512^3$ dark matter particles in a $(50 \, \mathrm{Mpc} \, h^{-1})^3$ box from $z = 99$ to $z = 0$. The four simulations are based on different realizations of the initial Gaussian random field, generated using \textsc{genetIC} \citep{Stopyra2021}. Three simulations were used for training the machine learning model and one was set aside for validation and testing. 

Dark matter halos were identified at $z=0$ using the SUBFIND halo finder \citep{gadget2, gadget4}, as done in Ref.~\cite{LucieSmith2022}. We restricted our analysis to halos within the mass range $\log_{10} ( M/M_{\odot} ) \in [11, 13]$, in order to fully resolve the inner profile of the lowest-mass halos and not be affected by small-number statistics at the high-mass end. To track the evolution history of the dark matter halos, we saved $91$ snapshots of the simulations between $z\sim7$ and $z=0$. We used the \textsc{pynbody} and \textsc{tangos} software packages \cite{Pontzen2018} to construct the merger trees of every dark matter halo. \textsc{tangos} matches a halo with its successor in time based on the fraction of common particles between the two objects; the procedure is repeated for every timestep in the simulation, thus yielding halo merger trees from $z\sim7$ to $z=0$. The merger trees were then used to track the mass of each halo's main progenitor over time.
\begin{figure}
      \centering
        \includegraphics[width=\columnwidth]{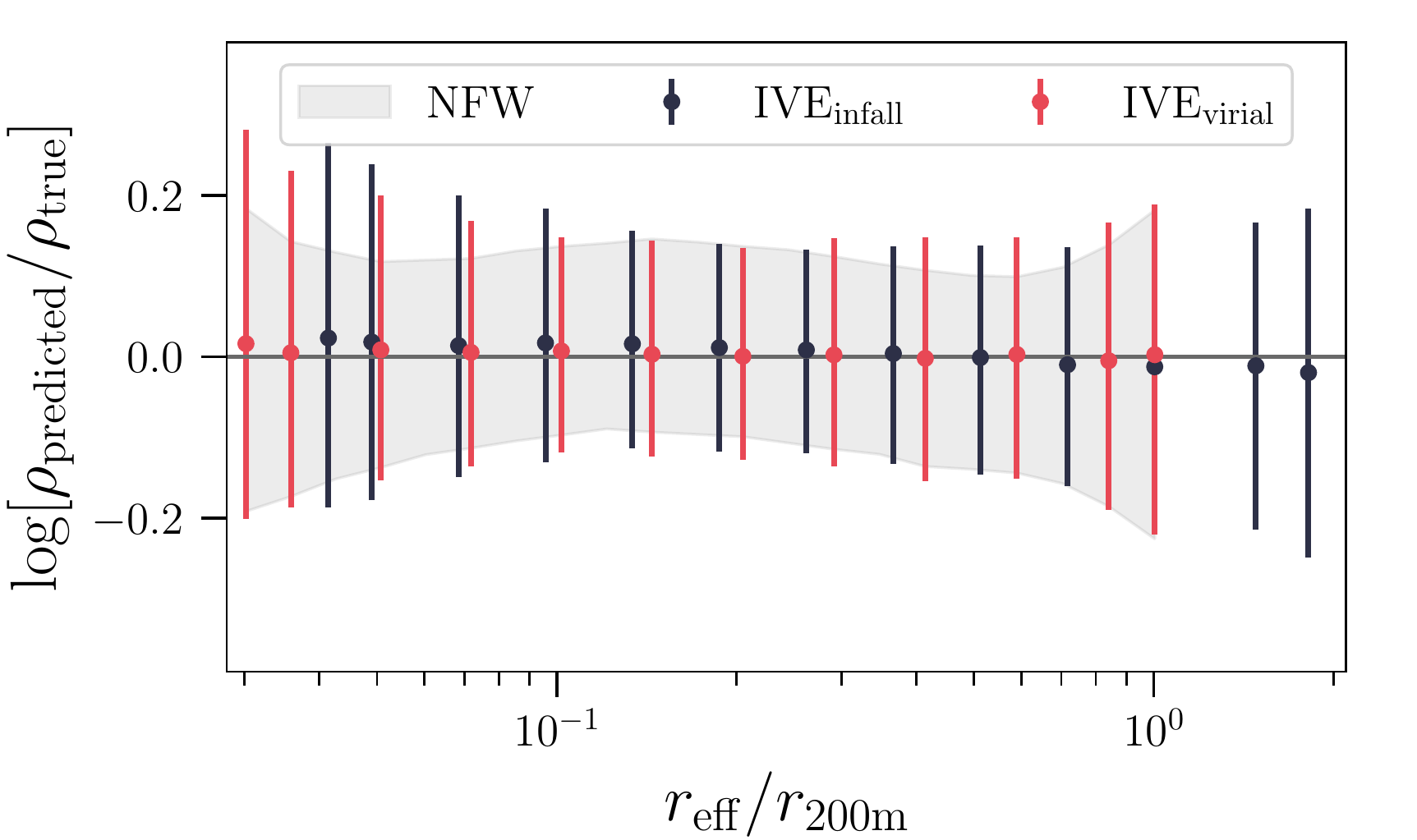}
        \caption{Mean and 90$\%$ confidence interval of the residuals $\log [\rho_{\rm predicted}/\rho_{\rm true}]$ of the \IVEi{} and \IVEo{} models, as a function of $r_{\rm eff}$ defined as the median radius in each bin. The grey band shows the NFW residuals.}
    \label{fig:predictions}
\end{figure}

\subsection{Predictions for the halo density profiles}

Figure~\ref{fig:predictions} shows the accuracy of the predictions of the \IVEi{} and \IVEo{} models. We show the mean and 90\% confidence interval of the residuals $\log_{10} [\rho_{\rm predicted}/\rho_{\rm true}]$, in every radial bin of the profile used for testing. Since every radial bin contains a different value of $r$ for different halos, we plot the residuals as a function of $r_{\rm eff}$ defined as the median of the distribution of radius values within each bin. The grey band shows the residuals of the NFW fits for comparison. Note that the IVE results include uncertainties in the latent distributions, whereas the NFW fits do not include uncertainties as they were obtained through least-squares optimization. The performance of the IVE models is consistent with that of the NFW profile, meaning that our model contains sufficient predictive accuracy to yield meaningful latent representations.

\subsection{Further physical interpretation of the inner shape latent}
  \begin{figure}
      \centering
        \includegraphics[width=\columnwidth]{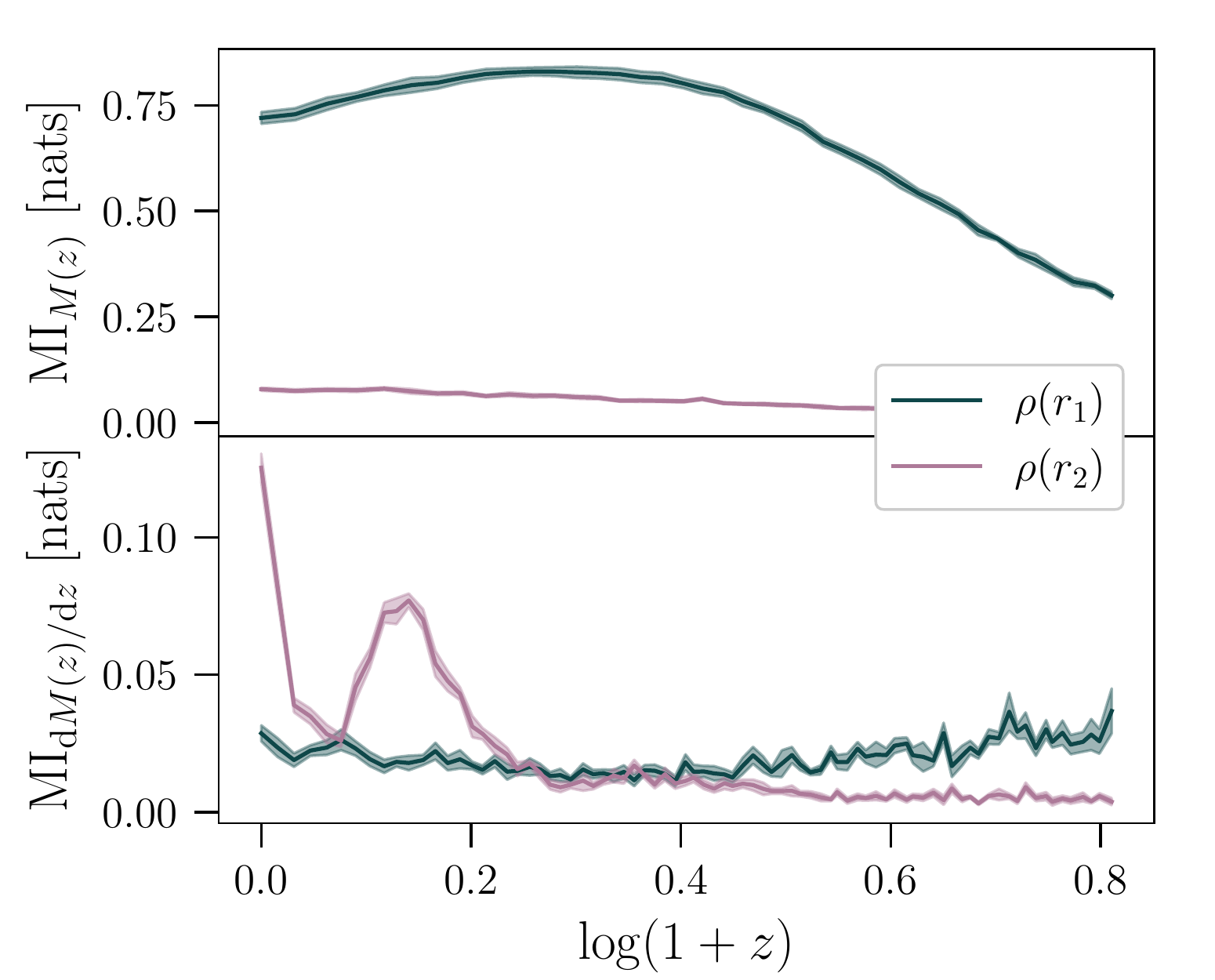}
        \caption{The MI between the densities on two fixed radial scales, $\rho(r_1)$ and $\rho(r_2)$, and (i) the mass accretion histories (top row) and (ii) the mass accretion rate (bottom row). The two radial scales, $r_1$ and $r_2$, are the locations at which the MI between the ground-truth profile and the inner shape latent peaks (Fig~\ref{fig:MI_latents_truth}, bottom panel).}
    \label{fig:MIrhopeaks_MAH}
\end{figure}

We present a further investigation on the physical interpretation of the inner shape latent. As shown in Fig.~\ref{fig:MI_latents_truth}, the MI between the inner shape latent and the ground-truth density profile has a bimodal shape.
The MI first peaks at $r_1 \sim 0.1 \, r_{\rm 200m}$ and then again at $r_2 \sim 0.9 \, r_{\rm 200m}$, meaning that the latent contains information about the shape of the profile in the inner region of the halo and close to the halo virial radius.
When physically interpreting the latent, we found that the latent carries information about the early formation history of the halo and the later time mass accretion rate (Fig.~\ref{fig:MI_latents_MAH}). We now verify if the early formation history is responsible for the shape of the profile in the inner region, while the later time mass accretion rate is what determines that closer to the virial radius. To do so, we compute the MI between the ground-truth density at the first MI peak, $\rho(r_1)$, and both the halo mass assembly history and mass accretion rate. We then repeat the calculation for the ground-truth density at the second MI peak, $\rho(r_2)$. Fig.~\ref{fig:MIrhopeaks_MAH} shows the MI between the two ground-truth densities with the mass assembly history, $M(z)$, in the top panel \footnote{Note that, in contrast to Fig.~\ref{fig:MI_latents_MAH}, here we consider the mass assembly history $M_{\rm 200m}(z)$ without normalizing it by the present-day halo mass. This is because $\rho(r)$ also depends on the overall normalization and therefore must be compared to $M_{\rm 200m}(z)$.} 
and that with the mass accretion rate, $\mathrm{d}\ln M(a)/\mathrm{d}\ln a$, in the bottom panel. The density in the inner region of the halo carries information about build up of mass from early times, but is largely uncorrelated with the mass accretion rate; the density close to the halo boundary is sensitive to the later time mass accretion rate, but does not depend on the early assembly phase. This confirms our physical interpretation of the bimodal shape of the MI between the latent and the profile in Fig.~\ref{fig:MI_latents_truth}: the way the inner profile responds to the latent depends on the early formation history of the halos, while the way the profile close to \rvir{} responds to that same latent depends on the later time mass accretion rate. The MI between $\rho(r_2)$ and the mass accretion rate shows an additional peak at $z\sim 0$ that is not as pronounced in the MI between the inner shape latent and the mass accretion rate. This is because, as one approaches the virial radius, the profile also becomes sensitive to the most recent mass accretion rate, which is captured by the outer latent.

\bibliography{paper}

\end{document}